\documentclass[12pt]{article}

\usepackage{color}
\usepackage{graphicx}
\usepackage{amsmath}
\usepackage{amssymb}
\usepackage{xspace}
\usepackage{hyperref}
\usepackage{indentfirst}
\usepackage{tikz}
\usepackage[sorting=none, maxbibnames=99, giveninits=true, style=numeric-comp]{biblatex}
\definecolor{PictureTitleColor}{rgb}{0.75,0,0} 
\definecolor{PictureAxesColor}{rgb}{0,0,0} 
\definecolor{PictureFormulaColor}{rgb}{0,0,0.50} 
\definecolor{PictureFlowColor}{rgb}{0,0,0.75} 
\definecolor{PictureTrajectoryColor}{rgb}{0.65,0.15,0.15} 
\definecolor{PicturePointColor}{rgb}{0,0,0.50} 

\definecolor{PictureGraphColor}{rgb}{0,0,0.75} 
\definecolor{PictureMarkingColor}{rgb}{0.65,0.15,0.15} 

\makeatletter
\@addtoreset{equation}{section}

\makeatletter
\renewcommand\section{\@startsection {section}{1}{\z@}%
             {-3.5ex \@plus -1ex \@minus -.2ex}
                                   {2.3ex \@plus.2ex}%
                                   {\normalfont\large\bfseries}}
\renewcommand\subsection{\@startsection{subsection}{2}{\z@}%
                    {-3.25ex\@plus -1ex \@minus -.2ex}%
                                     {1.5ex \@plus .2ex}%
                                     {\normalfont\bfseries}}

\def\baselinestretch{1.2}
\def\bear{\begin{eqnarray}}
\def\eear{\end{eqnarray}}
\def\nn{\nonumber}

\newcommand\secref[1]{{\S\ref{#1}}}
\newcommand\appref[1]{{Appendix~\ref{#1}}}

\newcommand\tabref[1]{{Table~\ref{#1}}}

\newcommand{\be}{\begin{equation}}
\newcommand{\ee}{\end{equation}}
\newcommand{\beq}{\begin{eqnarray}}
\newcommand{\eeq}{\end{eqnarray}}

\newcommand{\gone}[1]{{}}

\def\Ricci{{R}} 

\def\SizeL{{L}} 

\def\MPlanck{{M_p}} 

\def\AF{{A}} 
\def\AFBulk{{\mathcal A}} 
\def\FFBulk{{\mathcal F}} 

\def\PeriodT{{\tau}}
\def\FloquetH{{H_F}}

\def\bAF{{b}}

\def\CBogPlus{C_{+}} 
\def\CBogMinus{C_{-}} 
\def\wfM{{\widetilde{f}}} 
\def\aOp{{\mathbf{a}}} 
\def\Vol{{V}} 
\def\vp{{\vec{p}}} 
\newcommand\ket[1]{{\lvert{#1}\rangle}}
\newcommand\bra[1]{{\langle{#1}\rvert}}

\def\coB{{\mathbf{b}}}
\def\Ax{{A_x}}
\def\Ay{{A_y}}
\def\Jx{{J_x}}
\def\Jy{{J_y}}
\def\Ex{{E_x}}
\def\Ey{{E_y}}

\def\frr{f_{rr}}
\def\ftt{f_{tt}}
\def\fxx{f_{xx}}

\def\fyy{f_{yy}}

\def\unkB{{{\mathbf{A}}}}

\def\unkM{{{\mathbf{Z}}}}

\def\unkG{{{\mathbf{H}}}}
\def\unkX{{{\mathbf{X}}}}
\def\unkY{{{\mathbf{Y}}}}
\def\newr{{\rho}}

\def\pmtrC{{C}}
\def\coM{{C_\unkM}}
\def\coX{{C_\unkX}}
\def\coY{{C_\unkY}}
\def\newnewr{{u}}

\def\coB{{\beta}} 

\def\unkPhi{{\Phi}}

\def\ERapGrad{{\omega}}

\begin{document}
\begin{titlepage}

\vfil\

\begin{center}

{\Large{\bf
Timelike Kasner singularities and Floquet States in 2+1d AdS/CFT}}

\vfil

Emil Albrychiewicz and Ori J. Ganor

\vfil

Department of Physics, University of California,
Berkeley, CA 94720\\

Email: \href{mailto:ealbrych@berkeley.edu}{ealbrych@berkeley.edu}, \href{mailto:ganor@berkeley.edu}{ganor@berkeley.edu}

\vfil

\end{center}

\begin{abstract}
\noindent

We consider a model of a holographic 2+1d CFT interacting with an oscillating background gauge field. It is solved by an AdS-Vaidya metric describing Ohmic heating of the boundary field theory. However, we also show that if timelike singularities of Kasner type are permitted then a time independent solution that may be interpreted as a Floquet state of the system can be constructed. In this state the system exhibits either Hall conductivity or kinetic induction, and we numerically evaluate the Kasner exponents for a range of boundary conditions. This model may contribute to the ongoing discussion on the validity and meaning of the Kasner metric in the AdS/CFT correspondence and its application in cosmology.

\end{abstract}
\vspace{0.5in}

\end{titlepage}
\renewcommand{\baselinestretch}{1.05}  

\section{Introduction}
\label{sec:intro}

The Kasner metric with timelike singularity is
\be\label{eqn:KasnerMetric}
ds^2 = -u^{2p_t}dt^2+du^2+u^{2p_x}dx^2+u^{2p_y}dy^2,
\ee
where the range of $u$ is $u>0$, the singularity is at $u=0$, and $p_x,p_y,p_t$ are the constant {\it Kasner exponents}, which we will assume are all nonzero. If they satisfy
\be\label{eqn:KasnerCondition}
p_t+p_x+p_y=p_t^2+p_x^2+p_y^2=1
\ee
then \eqref{eqn:KasnerMetric} is an exact solution of Einstein's equations in vacuum \cite{Kasner:1921zz}. However, it is not clear if this kind of naked timelike singularity at $u=0$ is permitted in quantum gravity. There are reasons to suspect that solutions with naked Kasner singularities are forbidden in a holographic theory of quantum gravity. In \cite{Ren:2016xhb} Ren studied exact solutions of classical gravity with a negative cosmological constant that interpolate between AdS$_4$ near its boundary and a timelike Kasner singularity in the bulk, interpreting the solutions in the context of AdS/CFT \cite{Maldacena:1997re,Gubser:1998bc,Witten:1998qj}. Ren argued, using results of \cite{Iizuka:2012wt}, that the solutions do not satisfy the Gubser criterion \cite{Gubser:2000nd}, which posits that it should be possible to deform an admissible singularity to one hidden behind a horizon (thus allowing the introduction of a small positive temperature in the holographically dual QFT). And in \cite{Katoch:2023dfh}, Katoch et. al. computed the action of a ``Wheeler-DeWitt patch'', which is conjectured to be proportional to a ``Complexity'' measure of the holographically dual quantum state \cite{Susskind:2014rva,Brown:2015bva,Brown:2015lvg}. Katoch et. al. showed that the Complexity-Action is lower for the space with the naked Kasner singularity than for pure AdS, which they interpreted as an indication of the inadmissibility of the naked Kasner singularity. (See also \cite{Barbon:2015ria,Barbon:2015soa, Bolognesi:2018ion, Caputa:2021pad} for a related analysis of spacelike singularities.)
In contrast, the Kasner metric with a spacelike singularity is clearly viable, and plays a role in the interior of black holes, where the metric near the singularity can be decomposed into ``epochs'' at each of which the Kasner metric is a good description \cite{Frenkel:2020ysx, Hartnoll:2020fhc,Mansoori:2021wxf, Dias:2021afz, Grandi:2021ajl, Sword:2021pfm,Caceres:2022smh,Henneaux:2022ijt,An:2022lvo, Liu:2022rsy,Mirjalali:2022wrg}. 
Solutions with spacelike Kasner singularities in supergravity models have been recently described in \cite{Sabra:2022xmf}.

In this paper we explore a different aspect of the naked timelike Kasner singularity - we propose that if it is admissible, it would allow the construction of a ``trivial Floquet state'' in a CFT that is holographically dual to 3+1d Einstein-Maxwell gravity (with a negative cosmological constant as in AdS/CFT), irradiated by a uniform circularly polarized radiation. To be sure, what we mean by ``admissible'' is that there exists a system, which far below a certain UV scale, is describable by pure Einstein-Maxwell gravity with a negative cosmological constant, with a metric that approaches \eqref{eqn:KasnerMetric}. Near the UV scale, the system might have additional massive modes, which might even ``smooth out'' the metric.

 Floquet states are eigenstates of time translation by one whole period in a periodically driven system (see for instance \cite{Harper:2020,Viebahn:2020} for reviews). In any quantum system with a Hamiltonian that is periodic in time, with period $\PeriodT$, a Floquet state is an eigenstate of the time evolution operator $U(t_0,t_0+\PeriodT)$ from some specified time $t_0$ to $t_0+\PeriodT$. It is then, of course, automatically an eigenstate of all the evolution operators $U(t_0,t_0+n\PeriodT)$ by any integer number of periods $n$. Sometimes, a simple operator $\FloquetH$ can be identified such that $U(t_0,t_0+\PeriodT)=\exp\left(-i\PeriodT\FloquetH\right)$, and the operator is then referred to as a ``Floquet Hamiltonian''. Eigenstates of a Floquet Hamiltonian are Floquet states. The 2+1d CFT that we will study has a $U(1)$ current $J^\mu$. We introduce a periodicity by turning on a uniform background gauge field with a rotating electric field. A Floquet Hamiltonian for our system can be identified with the Hamiltonian in a rotating reference frame where the electric field is stationary. (Periodic quantum systems where the Floquet Hamiltonian can be derived by changing to a rotating reference frame where the system is stationary were referred to as ``trivial examples'' of Floquet systems in \cite{Viebahn:2020}. In our system we note, however, that if a discrete spectrum is desired we should formulate the system on a finite volume, with periodic boundary conditions, which would break rotational invariance and lead to a ``nontrivial'' Floquet system.)
 
 The minimal model we consider is a 2+1d CFT with a $U(1)$ conserved charge, dual to AdS$_4$ with only a metric and a gauge field. The bulk action takes the form
\begin{align}
\frac{1}{16\pi G_N}
\int \sqrt{-g}\left(\Ricci+\frac{6}{\SizeL^2}-\frac{1}{4}\FFBulk_{\mu\nu}\FFBulk^{\mu\nu}\right)d^4x,
\label{eqn:actionRF}
\end{align}
where $\FFBulk=d\AFBulk$ is the gauge field strength of the gauge field $\AFBulk$ that is holographically dual to the $U(1)$ current $J^\mu$ of the 2+1d CFT.
The pure Einstein-Maxwell theory is widely believed to be incomplete. Harlow and Ooguri \cite{Harlow:2018tng} argued that charged objects in the bulk are required for UV completeness, and an instability of pure AdS$_4$ was reported in \cite{Horowitz:2020tpa}. Nevertheless, the Einstein-Maxwell system is a reasonable toy model to exhibit the main point of our argument. If the introduction of massive fields resolves the Kasner singularity, with the masses typically in the order of the Planck scale $\MPlanck$, but the solution approximates a Kasner singularity at long range, then, by definition, the Kasner singularity is admissible in the classical limit.
In any case, we will not include additional fields in this paper.

A quantum field theory can respond to an oscillating background electromagnetic field in a variety of ways.
In a system with a finite dimensional Hilbert space, however, Floquet states are guaranteed to exist, since the time evolution operator over one period $U(t_0,t_0+\PeriodT)$ can be diagonalized. In a system with an infinite dimensional Hilbert space, while one can attempt to diagonalize $U(t_0,t_0+\PeriodT)$, the eigenstates are not guaranteed to have finite energy. A free field theory of charged particles, for example, absorbs radiation by converting photons into charged particle pairs \cite{DiPiazza:2011tq}. As the particle pairs are (composite) bosons, their presence amplifies the amplitude for absorption, and the system absorbs power at an ever growing rate, creating increasingly higher energy density. (We review the exact solution of this system in \secref{sec:FreeFields}.) Interacting quantum field theories typically have dissipative processes and resistivity, and indeed, the CFT model that we consider in this paper admits an AdS-Vaidya solution with Ohmic heating and energy density that is linearly increasing with time. (See \cite{Balasubramanian:2010ce} for an analysis of the thermalization process in AdS-Vaidya spacetimes.) In both cases the energy density increases indefinitely with time, but it is possible to design systems that avoid such an infinite thermalization \cite{DAlessio:2013,Ponte:2015}. In such systems one can find eigenstates of $U(t_0,t_0+\PeriodT)$ that have finite energy.

We proceed with a brief summary of previous works involving holography and Floquet behavior. In \cite{Auzzi:2013pca} the authors studied by perturbative means the properties of a strongly coupled CFT in a thermal state subject to deformations by a conformal operator that is periodic in time. Their work was further generalized for large amplitudes in \cite{Rangamani:2015sha}. Following the introduction of holographic superconductors \cite{Gubser:2008px,Hartnoll:2008vx}, driving by external fields was also introduced to those models. The 1+1d case was studied in \cite{Li:2013fhw,Natsuume:2013lfa} where a dynamical evolution of order parameters was discussed. The generalization for 2+1d space was shown in \cite{Ishii:2018ucz}, where the authors studied a Floquet state in a superconductor. Using holography, they studied the phase structure of Floquet states and described first and second order phase transitions. In \cite{Li:2020omw} non-equilibrium dynamics of holographic superconductors was addressed. 
The application of holography to Floquet states was also suggested in \cite{Hashimoto:2016ize, Kinoshita:2017uch}, where the authors studied both massless and massive D3-D7 flavor systems, obtained the phase structure and computed conductivities. Numerical values of critical exponents in such systems were computed \cite{Endo:2023vov}. The study of D3-D5 system was a subject of \cite{Garbayo:2020dmh,Berenguer:2022act}. The dynamics of a periodically driven complex scalar field was studied in \cite{Biasi:2017kkn}, and the possibility of driving with a gravitational field was discussed in \cite{Biasi:2019eap}. And in \cite{deBoer:2023lrd} a 1+1d CFT with a periodic perturbation constructed from Virasoro generators, proposed in \cite{Wen:2018agb}, was studied holographically in terms of deformations of the event horizon of a dual black hole. A driving setup in 3+1d, similar to ours, was discussed in \cite{Yang:2023dvk}, in the context of superfluids phase transition, in particular stable Floquet system was found in normal fluid phase. Our goal in the present paper, however, is not to construct a viable Floquet system, but rather to explore its relation to Kasner singularities, and therefore we settle for a minimalistic model.

The outline of this paper is as follows. In \secref{sec:FreeFields}, we review the behavior of a free charged scalar field interacting with an oscillating background field. Next, we provide a detailed description of our model in \secref{sec:ohmic}. In \secref{sec:Kasner}, we present our results for the Kasner exponents that appear when attempting to find a steady-state solution. We show numerically that the asymptotic solution of the equations of motion for our model leads to a Kasner metric. We follow in \secref{sec:discussion}, with a discussion of boundary conditions and conclusions. For completeness, in \appref{app:Vaidya} we include a derivation of the AdS-Vaidya metric in a general dynamic case. In \appref{app:MoreXYZorders}, we present higher order terms for a solution discussed in the main part. Finally, in \appref{app:KasnerExponents} we present numerical values of Kasner coefficients evaluated for a range of parameters.

\section{A free charged scalar field interacting with uniform radiation}
\label{sec:FreeFields}

To begin, we review the behavior of a free charged scalar QFT in 2+1d interacting with a background uniform radiation, which we take to be circularly polarized. (See \cite{DiPiazza:2011tq} for more details and references.) The free field QFT will serve as a benchmark for comparison to strongly interacting QFTs.

Our coordinates are $t,x,y$, and the scalar field $\phi(t,x,y)$ obeys the Klein-Gordon equation
\be\label{eqn:KG}
(\partial_\mu - i q A_\mu)(\partial^\mu - i q A^\mu)\phi=0,
\ee
where the index $\mu=0,1,2$ corresponds to $t,x,y$, the background gauge field $A_\mu$ is given by
\beq\label{eqn:AFdef}
\AF=A_\mu dx^\mu =\bAF(\cos\omega t\,dx+\sin\omega t\,dy),
\eeq
and $q$ is the charge. We will refer to \eqref{eqn:AFdef} as the ``laser beam''.
 As we will see, generally, the laser beam creates particle pairs of opposite momenta, in ranges of momentum bands that are optically transparent, while the remaining wavelengths can absorb the radiation. The location of the bands depends on the strength of the field (the constant $\bAF$), and is given by the sequence of bands between the characteristic numbers of Mathieu's equation, as we will review below.

One way to derive the ranges of momentum that can absorb radiation is to consider a more realistic setup whereby the field \eqref{eqn:AFdef} is not present for all $-\infty<t<\infty$ but is turned on at, say, $t=0$ and turned off at $t=T\gg 1/\omega$. To know which wavelengths continuously absorb radiation, we need to examine a solution 
\be\label{eqn:phipxpy}
\phi(t,x,y) = f(t)e^{i p_x x + i p_y y}
\ee
to the free field equations that has given momentum $(p_x,p_y)$, where the prefactor $f(t)$ behaves as $f(t)= e^{-i\Omega t}$ for $t<0$, where $\Omega=\sqrt{p_x^2+p_y^2}>0$ is a positive frequency. For $t>T$, after the field turns off again, the solution satisfies the free field equation of motion and can be decomposed into positive and negative frequencies:
$$
f(t) = \CBogPlus e^{-i\Omega t} + \CBogMinus e^{i\Omega t},
$$
where $\CBogPlus,\CBogPlus$ are Bogoliubov coefficients satisfying $|\CBogPlus|^2-|\CBogMinus|^2=1.$
Starting from the vacuum at time $t<0$, the average number of particle pairs with momentum $\pm p$ created from the vacuum at time $t=T$ is proportional to $|\CBogMinus|^2$, and if this increases exponentially with the time interval $T$, we have absorption of the laser beam.
In the case of a free boson, $|\CBogMinus|^2$  is related to the Floquet exponent $\mu$ of a Mathieu equation, as we will describe below.

Without loss of generality we can take $p_y=0$ and $p_x=p$ in \eqref{eqn:phipxpy}.
The Klein-Gordon equation \eqref{eqn:KG} becomes
\begin{align}
\label{eqn:Mathieu}
0 &=f''(t) + (p^2+q^2\bAF^2 - 2 p q\bAF\cos\omega t) f(t)
\end{align}
Before we discuss the actual solutions to \eqref{eqn:Mathieu}, let us describe how to derive the Floquet behavior from the solutions. As we assumed, our gauge field is turned on only for $0<t<T$, and we consider a solution $\wfM_{+}$ which satisfies \eqref{eqn:Mathieu} in the range $\epsilon<t<T-\epsilon$ [where $\epsilon$ is an unimportant time period that allows smooth interpolation from a zero field to \eqref{eqn:AFdef}] and satisfies the free field equation $\wfM_{+}'' + p^2\wfM_{+}=0$ outside the range $0<t<T$, with
$$
\wfM_{+}=e^{-i p t}\qquad
\text{for $t<0$.}
$$
For $t>T$, $\wfM_{+}$ again satisfies the free field equation, and therefore we must have
$$
\wfM(t) = \CBogPlus e^{-i p t} + \CBogMinus e^{i p t}\,,
$$
where $\CBogMinus$ and $\CBogPlus$ are Bogoliubov coefficients, and since \eqref{eqn:Mathieu} is real, the complex conjugate $\wfM_{-}\equiv(\wfM_{+})^\star$ is also a solution. It is then a standard observation that the Wronskian $\wfM_{+}'\wfM_{-}-\wfM_{-}'\wfM_{+}$ is constant, which implies
$$
|\CBogPlus|^2-|\CBogMinus|^2=1.
$$
We then expand the quantum field as
$$
\phi(t,x,y) = \int
\frac{dp_x dp_y}{\sqrt{2p\Vol}}
\Bigl\lbrack
\aOp_{\scriptstyle\text{early}}(\vp) e^{i (p_x x + p_y y)}\wfM_{+}(t;p)
+\aOp_{\scriptstyle\text{early}}(-\vp)^\dagger e^{i (p_x x + p_y y)}\wfM_{-}(t;p)
\Bigr\rbrack,
$$
where $\aOp_{\scriptstyle\text{early}}(\vp)$ are annihilation operators that annihilate the vacuum at early times ($t<0$), $\vp=(p_x,p_y)$ is the 2d momentum vector and $\Vol$ is a volume regulator. At late times ($t>T$), 
\begin{eqnarray}
\lefteqn{
\aOp_{\scriptstyle\text{early}}(\vp)\wfM_{+}(t;p)
+\aOp_{\scriptstyle\text{early}}(-\vp)^\dagger\wfM_{-}(t;p)
\longrightarrow
}\nonumber\\
&&
\Bigl(
\CBogPlus\aOp_{\scriptstyle\text{early}}(\vp)
+\CBogMinus^\star\aOp_{\scriptstyle\text{early}}(-\vp)^\dagger
\Bigr)e^{-i p t}
+\Bigl(
\CBogMinus\aOp_{\scriptstyle\text{early}}(\vp)
+\CBogPlus^\star\aOp_{\scriptstyle\text{early}}(-\vp)^\dagger
\Bigr)e^{i p t},
\nonumber
\end{eqnarray}
and so we identify the late-time annihilation operators (which annihilate the late-time vacuum) as
$$
\aOp_{\scriptstyle\text{late}}(\vp)\equiv
\CBogPlus\aOp_{\scriptstyle\text{early}}(\vp)
+\CBogMinus^\star\aOp_{\scriptstyle\text{early}}(-\vp)^\dagger.
$$
It now follows from the inverse transformation 
$$
\aOp_{\scriptstyle\text{early}}(\vp) = 
\CBogPlus\aOp_{\scriptstyle\text{late}}(\vp)
+\CBogMinus^\star\aOp_{\scriptstyle\text{late}}(-\vp)^\dagger,
$$
and the fact that the state (in the Heisenberg picture) is always annihilated by $\aOp_{\scriptstyle\text{early}}$, that the quantum state $\ket{\psi}$ satisfies
$$
\Bigl\lbrack
\CBogPlus\aOp_{\scriptstyle\text{late}}(\vp)
+\CBogMinus^\star\aOp_{\scriptstyle\text{late}}(-\vp)^\dagger
\Bigr\rbrack\ket{\psi}=0.
$$
Therefore, we see that we obtain a gas of particle pairs with opposite momenta $\vp$ and $-\vp$, and the average number of particles can easily be calculated:\footnote{
By comparing $\bra{\psi}\aOp_{\scriptstyle\text{late}}(\vp)^\dagger\Bigl\lbrack
\CBogPlus\aOp_{\scriptstyle\text{late}}(\vp)
+\CBogMinus^\star\aOp_{\scriptstyle\text{late}}(-\vp)^\dagger
\Bigr\rbrack\ket{\psi}=0$ to $\bra{\psi}\aOp_{\scriptstyle\text{late}}(-\vp)\Bigl\lbrack
\CBogPlus\aOp_{\scriptstyle\text{late}}(\vp)
+\CBogMinus^\star\aOp_{\scriptstyle\text{late}}(-\vp)^\dagger
\Bigr\rbrack\ket{\psi}=0$.
}
$$
\bra{\psi}
\aOp_{\scriptstyle\text{late}}(\vp)^\dagger
\aOp_{\scriptstyle\text{late}}(\vp)\ket{\psi}
=\bra{\psi}
\aOp_{\scriptstyle\text{late}}(-\vp)^\dagger
\aOp_{\scriptstyle\text{late}}(-\vp)
\ket{\psi}=
|\CBogMinus|^2,
$$
and we assume that $\CBogMinus$ and $\CBogPlus$ depend only on $p$ (but not on the direction), which is approximately true for large $T$ and small $\epsilon$.
(This derivation is reminiscent to the derivation of Hawking radiation \cite{Hawking:1975vcx}.)

Now we return to \eqref{eqn:Mathieu}. The solutions are Mathieu functions. 
By Floquet's theorem, there is a solution that satisfies 
$$
f\left(t+\frac{2\pi}{\omega}\right)=\sigma f(t)
$$
for some constant $\sigma(p,q)$ (the {\it Floquet coefficient}), and we would like to know for which values of $p$ (for given $q\bAF$) we have $|\sigma|>1$. For such values, there will be unbounded solutions $f(t)$, and $\CBogPlus$ will rise exponentially with $T$ [like $\exp(\frac{1}{2\pi}\omega T\log\sigma)$].
Whether such a $|\sigma|>1$ exists or not depends on the parameters $p$ and $q\bAF$.
In general, Mathieu's equation
$$
f''(t)+(\delta+\epsilon\cos t)f(t)=0
$$
exhibits this instability ($|\sigma|>1$) in certain known ranges of the parameters :
\begin{align}
\label{eqn:ParamMathieuRange}
\delta=\frac{p^2+q^2}{\omega^2},
\qquad
\epsilon=\frac{2pq}{\omega^2}\,.
\end{align}
The regions of instability in the $(\delta,\epsilon)$ plane are delineated by boundaries where there is (one) periodic solution.
These boundaries are denoted by a series of curves
$\delta=a_n(\epsilon)$ and $\delta=b_n(\epsilon)$ with
$$
a_n(0)=b_n(0)=\frac{n^2}{4}.
$$
(See \cite{kovacic2018mathieu} for a review.)
For $\epsilon\rightarrow 0$ almost all of the $0<\delta$-range is free of instabilities, but as $\epsilon$ gets larger, most of the $\delta$ space becomes unstable.
Thus, as $p\rightarrow 0$ we have stable solutions (unless $q=n/2$ exactly), but above some critical value of $p$ there will be unstable solutions.
As $p\rightarrow\infty$ there will be an infinite series of narrower and narrower strips in $p$-space where unstable solutions exist.

\begin{figure}[ht]
\includegraphics[width=\textwidth]{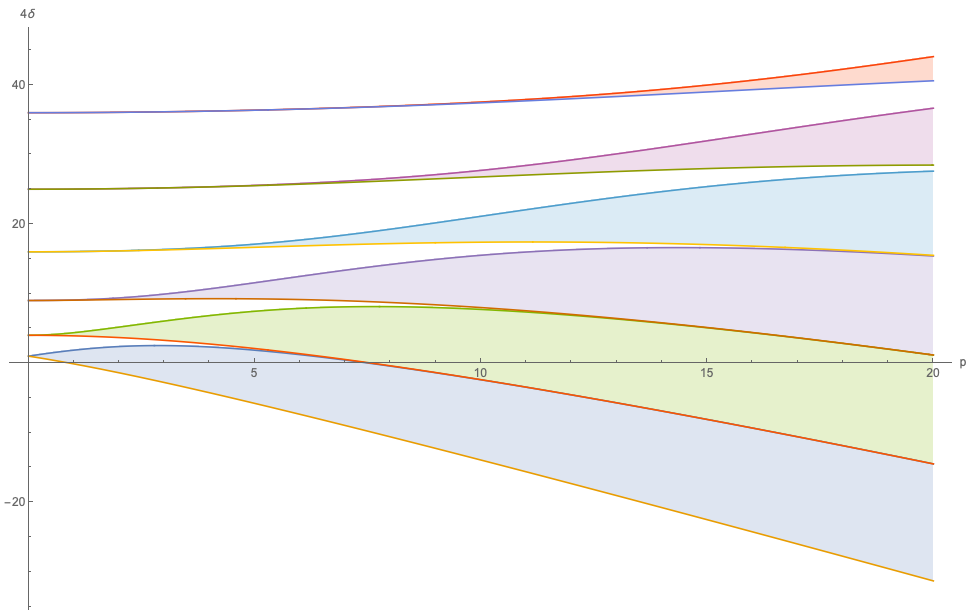}
\caption{A plot (using \texttt{Mathematica}) of the regions of $\delta,\epsilon$ \eqref{eqn:ParamMathieuRange} for which the solutions to Mathieu's equation are unbounded (hence signaling an instability). The range of momenta $p$ for which pairs will be created by the laser beam are the intersections of the graph of $\delta=p^2+q^2\bAF^2$ with the bands of instability. This plot is for $2q\bAF=1.5.$}
\end{figure}

There is a $p$ for which the Floquet coefficient $|\sigma(p,q)|$ is maximal, and it will generally be in the band closest to the IR (smaller $p$ values). We denote the momentum for which $|\sigma(p,q)|$ is maximal by $p_m$. Note that in general, $p_m\neq\omega$. This is a reflection of quantum corrections to the absorption frequency, and as $\bAF\rightarrow 0$, we find absorption only for $p$ being an integer multiple of $\omega$.

From the discussion above it follows that for any mass $m$, charge $q$, and amplitude $b$ the scalar field of our model absorbs energy from the laser beam at a rate that increases exponentially in time at a rate proportional to $\omega|\sigma(p,q)|^2$ for the value of $p$ for which this quantity is maximal, and the energy is converted to particle pairs with momentum $p$.
We will now turn to a strongly coupled holographic theory, which behaves in a qualitatively different way.

\section{The Holographic model with ohmic behavior}
\label{sec:ohmic}

Our model is a 2+1d CFT that is holographically dual to the Einstein-Maxwell 3+1d system described in \secref{sec:intro}, and we will analyze its behavior under a background field describing a uniform circularly polarized ``laser beam''. In other words, we need to solve the equations of motion that follow from \eqref{eqn:actionRF} with the gauge field $\mathcal{A
}_\mu$ approaching \eqref{eqn:AFdef} near the boundary, and the metric approaching the AdS$_4$ metric, as is standard in AdS/CFT \cite{Maldacena:1997re, Witten:1998qj, Gubser:1998bc}. The coordinates will be denoted by $r,t,x,y$, with $r$ corresponding to the bulk direction and $r=0$ is the boundary. (Not including time, our CFT is two-dimensional, and can be visualized, for example, as a thin film irradiated by a beam coming from a direction $z$ that is perpendicular to the sample, but such an auxiliary direction $z$ is irrelevant to our discussion below.) 
We also define a lightlike coordinate:
\be\label{eqn:lightcone}
v=t-r.
\ee
Practically, it is convenient to replace the time-independent amplitude $\bAF$ of the gauge field \eqref{eqn:AFdef} with a time-dependent amplitude that turns gradually from zero at $t\rightarrow -\infty$ to a finite value $\bAF$, say, at $t\ge 0$. We denote this amplitude by a function $\coB(t)$. The equations of motion then admit an AdS-Vaidya solution \cite{Vaidya:1951zz} that describes a bulk black hole with an expanding horizon, holographically dual to a process of Ohmic heating. The bulk gauge field of the solution is given by
\begin{align}
\label{eqn:AxAyExactSol}
\mathcal{A}_x=\coB(v)\cos\left(\omega v\right),\qquad
\mathcal{A}_y=\coB(v)\sin\left(\omega v\right),\qquad
\end{align}
and the metric in coordinates $x,y,v$ and a new coordinate $\eta$ (defined below) is
\begin{align}
\label{eqn:vaidya}
ds^2 &=-\left(\eta^2-\frac{f(v)}{\eta}\right)dv^2+ 2d\eta dv + \eta^2(dx^2+dy^2),
\end{align}
where the function $f$ is given by
\begin{align}
\label{eqn:fDef}
f(v)&=2\int_{-\infty}^v
\left\lbrack
\omega^2\coB(\tilde{v})^2+\coB'(\tilde{v})^2
\right\rbrack d\tilde{v}.
\end{align}
Note that we assume that $\coB(v)\rightarrow 0$ fast enough, as $v\rightarrow-\infty$, for the integral to converge.
In the metric \eqref{eqn:vaidya}, $\eta$ is a second lightlike coordinate that can be written as a function of $t,r$ using a solution of Abel's equation (see \appref{app:Vaidya} where we derive this metric for the general ansatz using Raychaudhuri equation \cite{Raychaudhuri}), but we will not need the details. The boundary is at $\eta=\infty$ and if $\coB$ vanishes the metric \eqref{eqn:vaidya} reduces to AdS$_4$ in Poincar\'e coordinates, $ds^2=(-dt^2+dr^2+dx^2+dy^2)/r^2$, after the change of variables $v=t-r$ and $\eta=1/r$.

The CFT current is derived from the $O(r)$ terms in the expansion of \eqref{eqn:AxAyExactSol} in powers of $r$ near the boundary:
\begin{align}
\label{eqn:JxJy}
G_N\Jx=-\omega\coB(t)\sin\left(\omega t\right)-\coB'(t)\cos\left(\omega t\right)
,\qquad
G_N\Jy=\omega\coB(t)\cos\left(\omega t\right)-\coB'(t)\sin\left(\omega t\right),
\end{align}
and after approaching steady state $\coB(t)$ can be replaced with the constant amplitude $\bAF$ so that
\begin{align}
\label{eqn:JxJySteady}
G_N\Jx=-\omega\bAF\sin\left(\omega t\right)=\Ex,\qquad
G_N\Jy=\omega\bAF\cos\left(\omega t\right)=\Ey,
\end{align}
where $(\Ex,\Ey)$ is the electric field derived from \eqref{eqn:AFdef}.
The Vaidya-AdS solution exhibits a linear Ohmic response of the system with conductivity
$$
\sigma = \frac{1}{G_N}.
$$
The solution \eqref{eqn:AxAyExactSol}-\eqref{eqn:vaidya} is a special case of a solution studied in \cite{Horowitz:2013mia}, where it was found that the conductivity is constant for any nonlinear driving field that depends only on the lightlike coordinate.\footnote{We are grateful to the JHEP referee for pointing this reference out to us.}

\section{Floquet behavior with Kasner singularities}
\label{sec:Kasner}

The Ohmic response described in \secref{sec:ohmic} is clearly and intuitively the correct behavior of the system when it starts from its ground state, however, after approaching its steady state (constant $\coB(t)=\bAF$) it breaks a symmetry that the equations of motion possess. Indeed, for constant $\bAF$, the background field \eqref{eqn:AFdef} is invariant not only under translations in the $x,y$ directions but also under a translation in time combined with a rotation in the $x,y$ plane, generated by the vector field
\be\label{eqn:Vfield}
\frac{\partial}{\partial t}-\omega y\frac{\partial}{\partial x}+\omega x\frac{\partial}{\partial y}\,.
\ee
If a uniform solution is also invariant under \eqref{eqn:Vfield} it must be invariant under time translations but for linear Ohmic heating the energy density grows linearly with time, and the system does not have an effective way to give energy back to the laser beam and reach a steady state with constant energy.
In this section, however, we explore the possibility that an admissible solution other than AdS-Vaidya exists and is invariant under \eqref{eqn:Vfield}. Such a solution will correspond to a Floquet state of the system.
We will also assume two additional discrete symmetries that are compatible with the driving field \eqref{eqn:AFdef}:
\begin{align}
\label{eqn:PT1}
t&\rightarrow -t,\qquad x\rightarrow x,\qquad y\rightarrow -y,\qquad r\rightarrow r,
\qquad
\AFBulk\rightarrow\AFBulk,
\end{align}
and
\begin{align}
\label{eqn:PT2}
t&\rightarrow t,\qquad x\rightarrow -x,\qquad y\rightarrow -y,\qquad r\rightarrow r,\qquad\AFBulk\rightarrow -\AFBulk.
\end{align}
The general form of an $x,y$ independent bulk metric with Killing vector \eqref{eqn:Vfield} and isometries \eqref{eqn:PT1} and \eqref{eqn:PT2} is
\begin{align} 
\label{eqn:BulkMetricAdS4}
ds^2&=\frr(r)dr^2-\ftt(r) dt^2+\fxx(r)(\cos\omega t\,dx+\sin\omega t\,dy)^2
\\ &
+\fyy(r)(\cos\omega t\,dy-\sin\omega t\,dx)^2.
\nonumber
\end{align}
where $\frr, \ftt,\fxx,\fyy$ are functions of $r$ only.
For the bulk gauge field we take the ansatz
\be\label{eqn:BulkGaugeField}
\AFBulk =\unkB(r)(\cos\omega t\,dx+\sin\omega t\,dy),
\qquad
\unkB(0)=\bAF.
\ee
We then choose the following parameterization for the functions that appear in \eqref{eqn:BulkMetricAdS4}:
\begin{align}
\label{eqn:VarDefKas}
\frr=\frac{1}{\unkG^2}e^{\unkX+\unkM}, \quad \fxx=e^{\frac{1}{2}(\unkX-\unkY)}, \quad \fyy=e^{\frac{1}{2}(\unkX+\unkY)}, \quad \ftt=e^{\unkM}\,.
\end{align}
the Lagrangian becomes
\begin{align}
    \label{eqn:LagStatCase}
    \mathcal{L}&= \frac{6e^{\unkX+\unkM}}{\unkG}
    +\frac{1}{2\unkG}\omega^2\unkB^2 e^{(\unkX-\unkY)/2}
    -\frac{1}{2}\unkG{\unkB'}^2 e^{-(\unkX-\unkY)/2}
    +\frac{\omega^2e^{\unkX}}{\unkG}(\cosh{\unkY}-1)
    \\ &
    +\frac{1}{2}\unkG\unkX'\unkM'
    +\frac{1}{8}\unkG\unkX'^2
    -\frac{1}{8}\unkG\unkY'^2\,.
\end{align}
We fix the parameterization freedom with the gauge choice $\unkG=1$, but keep the equation of motion that arises from varying $\unkG$:
\begin{align}
\label{eqn:EOMunkG}
0&=-6e^{\unkX+\unkM}
    -\tfrac{1}{2}\omega^2\unkB^2 e^{(\unkX-\unkY)/2}
    -\tfrac{1}{2}{\unkB'}^2 e^{-(\unkX-\unkY)/2}
    -\omega^2e^{\unkX}(\cosh{\unkY}-1)
    +\tfrac{1}{2}\unkX'\unkM'
    \\ \nonumber
    &+\tfrac{1}{8}\unkX'^2
    -\tfrac{1}{8}\unkY'^2\,.
\end{align}

The other (over determined) set of equations of motion is
\begin{align}
0=& \,
6e^{\unkX+\unkM}
+\tfrac{1}{4}\omega^2\unkB^2 e^{(\unkX-\unkY)/2}
+\tfrac{1}{4}{\unkB'}^2 e^{-(\unkX-\unkY)/2}
+\omega^2e^{\unkX}(\cosh{\unkY}-1) \label{eqn:EOM1} \\ 
& -\tfrac{1}{2}\unkM''
-\tfrac{1}{4}\unkX''
\,, \nonumber \\
0 =&
-\tfrac{1}{4}\omega^2\unkB^2 e^{(\unkX-\unkY)/2}
-\tfrac{1}{4}{\unkB'}^2 e^{-(\unkX-\unkY)/2}
+\omega^2e^{\unkX}\sinh\unkY
+\tfrac{1}{4}\unkY''
\,,\label{eqn:EOM2}\\
0 =&
\, 6e^{\unkX+\unkM}
-\tfrac{1}{2}\unkX''
\,,\label{eqn:EOM3}\\
0 =&
\left(\unkB' e^{-(\unkX-\unkY)/2}\right)'+
\omega^2\unkB e^{(\unkX-\unkY)/2}
\,.\label{eqn:EOM4}
\end{align}
Note that if we set the electromagnetic field to zero we get a solution
\be
\label{eqn:PureAdS}
\unkB=0,\qquad
\unkX=-\frac{4}{3}\log(3r),\qquad
\unkY=0,\qquad
\unkM=-\frac{2}{3}\log(3r).
\ee
This is a pure AdS$_4$ metric, which with the change of variables 
$$
r=\frac{\newr^3}{3},
$$
is brought to Poincar\'e form:
$$
ds^2 = \frac{-dt^2+d\newr^2+dx^2+dy^2}{\newr^2}\,.
$$
We expand around this AdS$_4$ solution, perturbatively in $\bAF$, and look for a solution in the form
\begin{align}
\unkX &= -4\log\newr + \sum_{n=1}^\infty\bAF^n\unkX_n
\,,\label{eqn:Xexpansion}\\
\unkY &= \sum_{n=1}^\infty\bAF^n\unkY_n
\,,\label{eqn:Yexpansion}\\
\unkM &= -2\log\newr+\sum_{n=1}^\infty\bAF^n\unkM_n
\,,\label{eqn:Mexpansion}\\
\unkB &= \sum_{n=1}^\infty\bAF^n\unkB_n
\,,\label{eqn:AbulkExp}
\end{align}
where the zeroth order corresponds to pure AdS$_4$. 
We find the following solution at first order
\begin{align}
\unkX_1 &= -\tfrac{1}{3}\pmtrC_3\newr^3
\,,\label{eqn:unkX1}\\
\unkY_1 &=
\tfrac{3}{8}\pmtrC_4(\sin2\newr-2\newr\cos 2\newr)
\,,\label{eqn:unkY1}\\
\unkM_1 &=\tfrac{1}{3}\pmtrC_3\newr^3
\,,\label{eqn:unkM1}\\
\unkB_1 &=
\pmtrC_1\cos\newr
+\pmtrC_2\sin\newr
\,,\label{eqn:unkB1}
\end{align}
where $\pmtrC_1,\dots,\pmtrC_4$ are constants.

In general, when perturbing around the AdS, solution the gauge field $\AFBulk$ has an expansion, in the gauge where $\AFBulk_r=0$,
\be\label{eqn:AbulkExpNearBdry}
\AFBulk_{i}(\newr,t,x,y)=\AF_{i}(t,x,y)+\newr J_{i}(t,x,y)+\cdots,
\ee
where $\AF_i$ is the background gauge field for the CFT and $J_i$ is the expectation value of the induced current \cite{Witten:1998qj,Gubser:1998bc}.
Therefore, the constants $\pmtrC_1, \pmtrC_2$, according to \eqref{eqn:AbulkExpNearBdry}, determine the gauge field source and the induced current at the boundary. We set $\pmtrC_1=1$ for the boundary condition on the gauge field to correspond to the field strength of the laser beam in \eqref{eqn:AFdef}.

The constants $\pmtrC_3, \pmtrC_4$ are related to the expectation values of components of the energy-momentum tensor,
\begin{align}
    \label{eqn:EnergyTensor}
    \langle T_{tt}\rangle&=\frac{1}{16\pi}\pmtrC_3, \qquad \langle T_{xx}\rangle=\frac{1}{32\pi}(-\pmtrC_3-3\pmtrC_4), \qquad \langle T_{yy}\rangle=\frac{1}{32\pi}(-\pmtrC_3+3\pmtrC_4).
\end{align}
In the standard AdS/CFT correspondence, with a black hole in the interior, the boundary energy density $\langle T_{tt}\rangle$ is a function of the location of the horizon in the bulk, and $\pmtrC_4=0$. These are determined by the condition of regularity in the interior. In our case, however, we have a Kasner singularity instead of a horizon in the bulk, and we are also allowing anisotropy in the $x-y$ directions. We therefore must allow, a priori, undetermined parameters $\pmtrC_3$ and $\pmtrC_4$. We will relate them to the Kasner parameters $p_t, p_x, p_y$ and the field strength $\bAF$, numerically in some special cases, later on. If, in the future, a better understanding of Kasner singularities emerges (if they are permissible at all), then $\pmtrC_3$ and $\pmtrC_4$ could be determined based on the allowed values of the Kasner exponents.

The expectation values \eqref{eqn:EnergyTensor} are derived, via the holographic dictionary, by a coordinate transformation that converts the metric near the boundary at $r=\newr=0$ to Fefferman-Graham (FG) form
\cite{Fefferman:2007rka,Graham:1999pm}:
\begin{align}
   \label{eqn:DefFefferman}
   \frac{1}{\newr^2}\left(d\newr^2+g_{ij}(\newr,x^k)dx^i dx^j\right),\qquad
    x^i= t,x,y. 
\end{align}
The required coordinate transformation is
\begin{align}
\label{eqn:FGTrans}
    \newr\rightarrow\newr+\xi(\newr, \bAF),
\end{align}
where the function $\xi$ is chosen so that the metric component $g_{\newr\newr}$ equals $\frac{1}{\newr^2}$, as in \eqref{eqn:DefFefferman}. It can be determined order by order in $\bAF$.
Following the procedure of holographic renormalization \cite{deHaro:2000vlm,Skenderis:2002wp}, the metric, written in FG coordinates \eqref{eqn:DefFefferman} on the boundary, has the asymptotic form
\begin{align}
\label{eqn:MetExp}
g_{ij}&=g^{(0)}_{ij}+\newr^2 g^{(2)}_{ij}+\newr^3 g^{(3)}_{ij}+O(\newr^4).
\end{align}
Then the expectation value of the boundary stress energy tensor is proportional to $g^{(3)}_{ij}$, and from \eqref{eqn:unkX1}-\eqref{eqn:unkM1} and the expansion \eqref{eqn:MetExp}, we get \eqref{eqn:EnergyTensor}.

When solving for $\unkX_n,\unkY_n,\unkM_n$ for $n\ge 2$, there are integration constants that allow adding a solution of the form \eqref{eqn:unkX1}-\eqref{eqn:unkM1} to $\unkX_n,\unkY_n,\unkM_n$. 
We can absorb these integration constants into the definition of the $\pmtrC_3$ and $\pmtrC_4$.
Then, the null energy condition on the boundary theory requires
\begin{align}
\label{eqn:NECcondition}    
-\frac{1}{3}\pmtrC_3\le\pmtrC_4\le\frac{1}{3}\pmtrC_3.
\end{align}

For $n>1$, $\unkX_n$ is a finite sum of the form
\begin{align}
\label{eqn:Xsol}
\unkX_n &= \sum_{k,m}\unkX_{n,k,m}\newr^k\cos m\newr
+\sum_{k,m}\widetilde{\unkX}_{n,k,m}\newr^k\sin m\newr
\end{align}
with constant coefficients $\unkX_{n,k,m}$ and $\widetilde{\unkX}_{n,k,m}$ that are zero unless
$$
\left\{
\begin{array}{lcll}
-\frac{3}{2}n\le k\le 3n & \text{and} & m\le 2n & \text{for even $n$,}\\
-\frac{3}{2}(n-1)\le k\le 3n-2 & \text{and} & m\le 2n-2 & \text{for odd $n$.}\\
\end{array}
\right.
$$
A similar formula holds for $\unkM_n$, expressed in terms of constant coefficients $\unkM_{n,k,m}$ and $\widetilde{\unkM}_{n,k,m}$ that satisfy the same conditions as above. 
As for the $\unkY_n$'s, we have
\begin{align}
\label{eqn:Ysol}
\unkY_n = \sum_{k,m}\unkY_{n,k,m}\newr^k\cos m\newr
+\sum_{k,m}\widetilde{\unkY}_{n,k,m}\newr^k\sin m\newr,
\end{align}
with coefficients $\unkY_{n,k,m}$ and $\widetilde{\unkY}_{n,k,m}$ that are zero unless
$$
\left\{
\begin{array}{lcll}
-\frac{3}{2}n\le k\le 3n & \text{and} & m\le 2n-2 & \text{for even $n$,}\\
-\frac{3}{2}(n-1)\le k\le 3n-1 & \text{and} & m\le 2n & \text{for odd $n$.}\\
\end{array}
\right.
$$
Explicit expressions for $\unkX_2,\unkY_2$ and $\unkM_2$ are listed in \appref{app:MoreXYZorders}

We do not know of an exact solution for the equations of motion \eqref{eqn:EOMunkG}-\eqref{eqn:EOM4}, and so we proceed numerically. We took the perturbative solutions, which we calculated up to $20^{th}$ order in $\bAF$, to specify initial conditions at some small $\newr=\newr_0>0$, and then we solved numerically for the region $\newr>\newr_0$. Asymptotically, for $\newr\rightarrow\infty$ we find that $\unkB$ reaches a constant, $\unkB\rightarrow\unkB_\infty$, and $\unkX,\unkY,\unkM$ generally behave as
\begin{align}
\unkX\sim-\tfrac{2}{3}(\coM+\coX)\newr^3,
\qquad
\unkY\sim\tfrac{2}{3}\coY\newr^3,
\qquad
\unkM\sim\tfrac{2}{3}\coM\newr^3,
\end{align}
where $\coX,\coY,\coM$ are constants that depend on $\pmtrC_1,\dots,\pmtrC_4$. We can now change our initial metric variables to $\newnewr$ given by
\begin{align}
\newnewr&=e^{-\coX r}=e^{-\frac{1}{3}\coX\newr^3},
\end{align}
with the range $0\leq\newnewr\leq 1$. (The AdS boundary is located at $\newnewr=1$.)
The change of variables shows that the asymptotic form of the metric as $\newr\rightarrow\infty$ ($\newnewr\rightarrow 0$) is of Kasner type \eqref{eqn:KasnerMetric} with Kasner exponents given by
\begin{align}
p_t = -\frac{\coM}{\coX}\,,
\qquad
p_x = \frac{\coX+\coY+\coM}{2\coX}\,,
\qquad
p_y = \frac{\coX-\coY+\coM}{2\coX}\,.
\label{eqn:ptpxpy}
\end{align}
We report in \appref{app:KasnerExponents} the numerical values that we found for the Kasner exponents $p_t, p_x, p_y$ for various values of the parameters $\pmtrC_2,\pmtrC_3,\pmtrC_4$. The latter are related to induced current ($\pmtrC_2$) and energy momentum tensor of the boundary CFT, as in \eqref{eqn:EnergyTensor}.

\section{Discussion}
\label{sec:discussion}

We have explored the possibility that a singularity of the form \eqref{eqn:KasnerMetric} is permissible within the framework of the AdS$_4$/CFT$_3$ correspondence, and argued that it would allow the construction of an excited state of the holographically dual QFT with a Floquet behavior. 
The Kasner solution we proposed, which describes a steady state, corresponds to a vanishing power flow from the background electromagnetic field to the CFT, balancing emission and absorption by the material.
This corresponds to either kinetic induction or Hall conductivity, as we shall now argue. 
We cannot distinguish between those, because we do not have a good understanding of the boundary conditions for perturbations of the metric and gauge field near the Kasner singularity, if it is viable at all. (I.e., if there exists a concrete model that reduces to Einstein-Maxwell gravity at low energy in the bulk and where the timelike Kasner singularity is ``resolved'' at the Planck scale.)

We can, however, consider the relationship between the current and gauge field \eqref{eqn:AbulkExpNearBdry} in the given solution \eqref{eqn:unkB1}.
The first case we can consider is where the material is a perfect insulator and the current vanishes with $\pmtrC_2=0$. This corresponds to Neumann boundary conditions for $\unkB_1$, where the value of a normal component of the field strength is fixed to zero. The system is also non-dissipative as Joule heating vanishes. (In a more realistic material, once we exceed a critical value of the electric field, the Schwinger effect should result in pair production and their motion induces a boundary current. Clearly, in such a condition there will be dissipation and our assumption will no longer hold.)

On the other hand, when $\pmtrC_2\neq 0$, using the expansion \eqref{eqn:AbulkExpNearBdry} and the solution \eqref{eqn:unkB1}, the boundary current satisfies
\begin{align}
    \label{eqn:BdryCurrent}
    \mathcal{E}_i=\mathcal{B}\frac{d\mathcal{J}_i}{dt}+\mathcal{C}\epsilon_{ij}\mathcal{J}_j,
\end{align}
where $\epsilon_{ij}$ is the Levi-Civita symbol. The first term of \eqref{eqn:BdryCurrent} is the first London equation and the second term is Hall conductivity. We introduced unknown coefficients $\mathcal{B}$ and $\mathcal{C}$, and the boundary conditions only constrain their sum, but not their difference. Therefore, we cannot distinguish kinetic induction for Hall conductivity without a better understanding of the tentative Planck scale resolution of the Kasner singularity.

The  London term in \eqref{eqn:BdryCurrent} would indicate that the boundary material is in a superconducting phase. In the context of holographic Floquet models superconductivity has appeared before in \cite{Li:2013fhw, Natsuume:2013lfa, Ishii:2018ucz, Li:2020omw}. In \cite{Ishii:2018ucz}, phase transitions were studied in the system of a holographic superconductor with periodic driving, and it was found that a rotating external field does not enhance superconductivity (rather it lowers the transition temperature). However, for low temperatures the normal current stayed suppressed and by considering the modified free energy it was argued that one obtains a steady state. 

From \eqref{eqn:BdryCurrent}, in the general case, we also have a non-zero transverse conductivity. This is a characteristic of the Hall effect. Although in our model we do not introduce an external magnetic field, both parity and time reversal symmetry are broken. 
It is noteworthy that in \cite{Garbayo:2020dmh} it was shown that Floquet driving can enhance the appearance of the quantum Hall effect. For a gauge field theory, which does not have to be periodic, in the bulk a topological term $\theta$ results in Chern-Simons action on the boundary \cite{Witten:2003ya}. This action serves as a source for current two-point functions that are proportional to the antisymmetric Levi-Civita symbol leading to Hall conductivity \cite{Hartnoll:2007ai, Keski-Vakkuri:2008ffv}.  

It would be interesting to better understand the stability of the stationary solutions described in this paper, especially in light of recent work on the instability of timelike Kasner-AdS metrics. As mentioned before, in \cite{Katoch:2023} the authors evaluated the Complexity Action for Kasner-AdS geometry and compared it to the empty Poincar\'e solution. In that way, they showed that a particular range of Kasner exponents is allowed, and those agree with the general Gubser criterion \cite{Gubser:2000nd}. If such a method was admissible in our model it would provide further restrictions on the $\pmtrC_i$ coefficients [in addition to NEC \eqref{eqn:NECcondition}], but unfortunately the limitations of our numerical techniques make it difficult to explore at the moment. One can question whether the appearance of the Kasner singularity indicates the limit of validity of the AdS/CFT correspondence. It is widely believed that the holographic dictionary should hold as long as a causality constraint \cite{Kleban:2001nh} is satisfied. And as was shown in \cite{Ren:2016xhb}, for the Kasner-AdS solution, null geodesics connecting two boundary worldlines at two different $(x,y)$ values are extremal (in the sense of minimal $\Delta t$) at the boundary, regardless of the value of the Kasner exponents. 

The calculations reported in this paper can be interpreted as follows: if a timelike Kasner singularity is permissible, presumably for some particular UV completion of the Einstein-Maxwell theory, a (highly excited) state of the CFT dual can be prepared in an unusual Floquet state, where the induced boundary current is not governed by Ohmic resistance, as it is in the Vaidya solution.

\section*{Acknowledgments}
We wish to thank Andr\'es Franco Valiente and Chao Ju for illuminating discussions. This work has been supported by the Berkeley Center for Theoretical Physics.


\begin{appendix}

\section{AdS-Vaidya metric and Abel equation}
\label{app:Vaidya}
In this section let us briefly show how to derive the metric \eqref{eqn:vaidya}, using the action \eqref{eqn:actionRF} and the metric ansatz:
\begin{align}
\label{eqn:metricMFK}
ds^2 = e^{\unkM}(dr^2-dt^2)
+e^{\frac{1}{2}\unkX}\left(
dx^2+dy^2
\right),
\end{align}
where $\unkM, \unkX$ are now functions of $r,t,$ and we fixed a gauge choice ($A_r=0$) as well as parameterization invariance. 
The Lagrangian takes the form
\begin{align}
\label{eqn:DynLagNoWY}
\mathcal{L} &=
6e^{\unkM + \frac{1}{2}\unkX} 
+\tfrac{1}{2}(\partial_t\Ax)^2
+\tfrac{1}{2}(\partial_t\Ay)^2 
-\tfrac{1}{2}(\partial_r\Ax)^2
-\tfrac{1}{2}(\partial_r\Ay)^2 
\\ &
\nonumber
-\tfrac{1}{2}e^{\frac{1}{2}\unkX}\partial_t\unkM\partial_t\unkX 
-\tfrac{1}{8}e^{\frac{1}{2}\unkX}(\partial_t\unkX)^2
+\tfrac{1}{2}e^{\frac{1}{2}\unkX}\partial_r\unkM\partial_r\unkX 
+\tfrac{1}{8}e^{\frac{1}{2}\unkX}(\partial_r\unkX)^2.
\end{align}
The main advantage of the gauge choice \eqref{eqn:metricMFK} is that the equation of motion for the gauge field simplifies to the free wave equation in $r,t$ variables, assuming $\Ax,\Ay$ depend on $r,t$ only. Hence, we can set the exact solution, in the retarded form required for causality as in \eqref{eqn:AxAyExactSol}, which we repeat here for convenience:
\begin{align}
\label{eqn:AxAyExactSol1}
\Ax=\coB(v)\cos\left(\ERapGrad v\right),\qquad
\Ay=\coB(v)\sin\left(\ERapGrad v\right),
\end{align}
recalling that $v$ is the lightcone coordinate: $v=t-r.$

Inspired by Raychaudhuri's equation, for a family of null geodesics given by constant $v,x,y$, we can combine the equations of motion, which come from a variation of the Lagrangian \eqref{eqn:DynLagNoWY}, to yield
\begin{align}
    \label{eqn:Raychaudhuri}
    \partial_u\left(e^{-\unkM}\partial_u\unkX\right)
    =-\frac{1}{4}e^{-\unkM}(\partial_u\unkX)^2,
\end{align}
where the second lightcone coordinate $u=t+r$ was used.
Defining
$$
\unkPhi = e^{-\frac{1}{4}\unkX},
$$
we can further simplify \eqref{eqn:Raychaudhuri} using the equation of motion [derived from \eqref{eqn:DynLagNoWY}] and the boundary conditions to get
\begin{align}
\label{eqn:Abel}
\frac{\partial\unkPhi}{\partial v} = \coB^2 v\unkPhi^3 + 1,
\end{align}
which is known as an {\it Abel equation of the first kind.}\footnote{
Abel's equation appears in other, unrelated, context in General Relativity. For example, in \cite{Yurov:2008sy} the authors express the solutions to Friedmann-Lemaitre-Robertson-Walker cosmology in terms of a solution to a suitable Abel equation.}
There are only several special cases in which analytic solutions to this equation are known. The most comprehensive list we are aware of is contained in \cite{polyanin2017handbook}. Unfortunately, our equation does not belong to any of the classes mentioned therein. However, we do not need to solve this equation. Instead we can use it to rewrite the metric \eqref{eqn:metricMFK} as
\begin{align}
\label{eqn:VaidyaMetMT}
ds^2 &=-\left(\eta^2-\frac{f(v)}{\eta}\right)dv^2+ 2d\eta dv + \eta^2(dx^2+dy^2),
\end{align}
and to recognize that it is an AdS$_4$-Vaidya metric 
with $\eta$ coordinate determined by the solution for $\unkPhi$, with the coordinate redefinition
$$
\eta=\frac{1}{\unkPhi}.
$$
The boundary is now at $\eta=\infty$ and the gauge-field-dependent function $f$ is given by
$$
f(v)=2\int_{-\infty}^v
\left\lbrack
\ERapGrad^2\coB(\tilde{v})^2+\coB'(\tilde{v})^2
\right\rbrack d\tilde{v}.
$$
Thus,  using solely the action and the general ansatz we re-derive the AdS-Vaidya solution for the bulk of this holographic model with oscillatory gauge field driving.


\section{Higher orders of $\unkX_n$, $\unkY_n$ and $\unkM_n$}
\label{app:MoreXYZorders}

Below we list second order solutions for the system of equations \eqref{eqn:EOMunkG}-\eqref{eqn:EOM4}:
\begin{align}
    \unkX_2 =\,&
    -\frac{3}{28}\pmtrC_1^2\newr^4
    -\frac{3}{28}\pmtrC_2^2\newr^4 
    +\frac{9}{2048}\pmtrC_4^2
    +\frac{27}{1280}\pmtrC_4^2\newr^2 
    -\frac{27}{448}\pmtrC_4^2\newr^4
    -\frac{1}{72}\pmtrC_3^2\newr^6
    \nn\\ &
    +\frac{27}{2048}\pmtrC_4^2\cos4\newr
    -\frac{27}{4096\newr^2}\pmtrC_4^2\cos4\newr
    +\frac{27}{16384\newr^3}\pmtrC_4^2\sin4\newr
    -\frac{27}{2048\newr}\pmtrC_4^2\sin4\newr,
    \nn \\
    \unkY_2 =\,&
    \frac{1}{8}(\pmtrC_1^2+\pmtrC_2^2)
    +\frac{1}{4}(\pmtrC_1^2+\pmtrC_2^2)\newr^2
    -\frac{1}{8}(\pmtrC_1^2+\pmtrC_2^2)\cos 2\newr
    -\frac{1}{4}(\pmtrC_1^2+\pmtrC_2^2)\newr\sin 2\newr
    \nn\\ &
    +\frac{3}{64}\pmtrC_4\pmtrC_3\newr^3\sin 2\newr
    -\frac{3}{32}\pmtrC_4\pmtrC_3\newr^4\cos 2\newr
    -\frac{1}{16}\pmtrC_4\pmtrC_3\newr^5\sin 2\newr,
    \nn \\
    \unkM_2 =\,&
    \frac{1}{14}(\pmtrC_1^2+\pmtrC_2^2)\newr^4
    -\frac{9}{2048}\pmtrC_4^2
    -\frac{63}{2560}\pmtrC_4^2\newr^2
    -\frac{9}{224}\pmtrC_4^2\newr^4
    -\frac{1}{144}\pmtrC_3^2\newr^6
    \nn\\ &
    +\frac{27}{2048}\pmtrC_4^2\cos 4\newr
    -\frac{27}{8192\newr^2}\pmtrC_4^2\cos 4\newr
    -\frac{9}{512\newr^2}\pmtrC_4^2\newr^2\cos 4\newr
    \nn\\ &
    -\frac{27}{32768\newr^3}\pmtrC_4^2\sin 4\newr
    -\frac{27}{4096\newr}\pmtrC_4^2\sin 4\newr
    +\frac{27}{1024\newr}\pmtrC_4^2\newr\sin 4\newr.
    \nn
\end{align}

\section{Numerical results for Kasner exponents}
\label{app:KasnerExponents}

\tabref{fig:TableKasner} and \tabref{fig:TableKasner2} show results for the Kasner exponent $p_t$ of \eqref{eqn:ptpxpy}, given a sample choice of parameters $C_2$ [which determines the induced boundary current via \eqref{eqn:unkB1} and \eqref{eqn:AbulkExpNearBdry}] and $C_3, C_4$ [which determine the boundary stress energy tensor \eqref{eqn:EnergyTensor}].
They were evaluated by a numerical solution starting at $\newr=1$ with initial conditions calculated from the perturbative solution in $\bAF$ (which we picked as $\bAF=0.25$) using \texttt{Mathematica}. For \tabref{fig:TableKasner}, $\pmtrC_2$ vanishes indicating the case of insulator on the boundary. For \tabref{fig:TableKasner2}, we turn on current by setting $\pmtrC_2=1$. In the range of tested parameters we include values for which classical energy condition \eqref{eqn:NECcondition} is broken. 

\begin{table}
\centering
\resizebox{\columnwidth}{!}{%
\begin{tabular}{|l|l|l|l|l|l|l|l|l|l|l|l|}
\cline{2-12} 
\multicolumn{1}{c}{}&
\multicolumn{11}{|c|}{
{\bf Numerical values of $p_t$ for $\bAF=0.25, \pmtrC_1=1, \pmtrC_2=0$}}
\\
\cline{2-12}
\multicolumn{1}{c}{}&
 \multicolumn{11}{|c|}{$\pmtrC_4$}
 \\
 \hline
$\pmtrC_3\downarrow$ &
{\bf -0.5} &
{\bf -0.4} &
{\bf -0.3} &
{\bf -0.2} &
{\bf -0.1} &
{\bf 0.0} &
{\bf 0.1} &
{\bf 0.2} &
{\bf 0.3} &
{\bf 0.4} &
{\bf 0.5} 
\\
\hline
{\bf -0.5} & 
-0.333 & 
-0.333 & 
-0.318 & 
-0.303 & 
-0.294 & 
-0.290 & 
-0.288 & 
-0.281 & 
-0.267 & 
-0.248 & 
-0.241  
\\
\hline
{\bf -0.4} & 
-0.301 & 
-0.317 & 
-0.322 & 
-0.322 & 
-0.318 & 
-0.312 & 
-0.307 & 
-0.305 & 
-0.311 & 
-0.322 & 
-0.331  
\\
\hline
{\bf -0.3} & 
-0.263 & 
-0.266 & 
-0.277 & 
-0.290 & 
-0.301 & 
-0.311 & 
-0.319 & 
-0.325 & 
-0.329 & 
-0.331 & 
-0.331  
\\
\hline
{\bf -0.2} & 
-0.326 & 
-0.319 & 
-0.312 & 
-0.308 & 
-0.307 & 
-0.307 & 
-0.308 & 
-0.308 & 
-0.306 & 
-0.301 & 
-0.293  
\\
\hline
{\bf -0.1} & 
-0.333 & 
-0.333 & 
-0.331 & 
-0.327 & 
-0.322 & 
-0.317 & 
-0.311 & 
-0.304 & 
-0.297 & 
-0.290 & 
-0.284  
\\
\hline
{\bf 0.0} & 
-0.333 & 
-0.333 & 
-0.332 & 
-0.330 & 
-0.328 & 
-0.325 & 
-0.321 & 
-0.318 & 
-0.314 & 
-0.312 & 
-0.312  
\\
\hline
{\bf 0.1} & 
-0.330 & 
-0.330 & 
-0.330 & 
-0.329 & 
-0.329 & 
-0.328 & 
-0.327 & 
-0.327 & 
-0.327  & 
-0.327 & 
-0.328  
\\
\hline
{\bf 0.2} & 
-0.324 & 
-0.326 & 
-0.327 & 
-0.328 & 
-0.329 & 
-0.329 & 
-0.330 & 
-0.331 & 
-0.332 & 
-0.332 & 
-0.333  
\\
\hline
{\bf 0.3} & 
-0.320 & 
-0.322 & 
-0.324 & 
-0.327 & 
-0.328 & 
-0.330 & 
-0.331 & 
-0.332 & 
-0.333 & 
-0.333 & 
-0.333  
\\
\hline
{\bf 0.4} & 
-0.317 & 
-0.320 & 
-0.323 & 
-0.326 & 
-0.328 & 
-0.330 & 
-0.332 & 
-0.333 & 
-0.333 & 
-0.333 & 
-0.333  
\\
\hline
{\bf 0.5} & 
-0.317 & 
-0.320 & 
-0.323 & 
-0.326 & 
-0.329 & 
-0.330 & 
-0.332 & 
-0.333 & 
-0.333 & 
-0.333 & 
-0.333  
\\
\hline
\end{tabular}%
}
\caption{Numerical values of Kasner exponent $p_t$, evaluated numerically for an insulator case.}
\label{fig:TableKasner}
\end{table}

\begin{table}
\centering
\resizebox{\columnwidth}{!}{%
\begin{tabular}{|l|l|l|l|l|l|l|l|l|l|l|l|}
\cline{2-12} 
\multicolumn{1}{c}{}&
\multicolumn{11}{|c|}{
{\bf Numerical values of $p_t$ for $\bAF=0.25, \pmtrC_1=1, \pmtrC_2=1$}}
\\
\cline{2-12}
\multicolumn{1}{c}{}&
 \multicolumn{11}{|c|}{$\pmtrC_4$}
 \\
 \hline
$\pmtrC_3\downarrow$ &
{\bf -0.5} &
{\bf -0.4} &
{\bf -0.3} &
{\bf -0.2} &
{\bf -0.1} &
{\bf 0.0} &
{\bf 0.1} &
{\bf 0.2} &
{\bf 0.3} &
{\bf 0.4} &
{\bf 0.5} 
\\
\hline
{\bf -0.5} & 
-0.304 & -0.327 & -0.332 & -0.331 & -0.328 & -0.322 & -0.317 & -0.313 \
& -0.311 & -0.311 & -0.311 
\\
\hline
{\bf -0.4} & 
-0.296 & -0.322 & -0.330 & -0.330 & -0.327 & -0.323 & -0.319 & -0.316 \
& -0.315 & -0.314 & -0.315 
\\
\hline
{\bf -0.3} & 
-0.291 & -0.316 & -0.326 & -0.328 & -0.327 & -0.324 & -0.321 & -0.319 \
& -0.318 & -0.318 & -0.318 
\\
\hline
{\bf -0.2} & 
-0.289 & -0.310 & -0.322 & -0.326 & -0.326 & -0.324 & -0.323 & -0.322 \
& -0.321 & -0.321 & -0.321 
\\
\hline
{\bf -0.1} & 
-0.289 & -0.304 & -0.317 & -0.323 & -0.325 & -0.325 & -0.324 & -0.324 \
& -0.324 & -0.324 & -0.324   
\\
\hline
{\bf 0.0} & 
-0.290 & -0.299 & -0.311 & -0.319 & -0.323 & -0.325 & -0.326 & -0.326 \
& -0.326 & -0.326 & -0.326 
\\
\hline
{\bf 0.1} & 
-0.291 & -0.293 & -0.305 & -0.316 & -0.323 & -0.326 & -0.327 & -0.328 \
& -0.328 & -0.328 & -0.328 
\\
\hline
{\bf 0.2} & 
-0.292 & -0.287 & -0.299 & -0.313 & -0.322 & -0.327 & -0.329 & -0.330 \
& -0.330 & -0.330 & -0.330 
\\
\hline
{\bf 0.3} & 
-0.291 & -0.281 & -0.294 & -0.311 & -0.322 & -0.327 & -0.330 & -0.331 \
& -0.331 & -0.332 & -0.331 
\\
\hline
{\bf 0.4} & 
-0.288 & -0.274 & -0.288 & -0.309 & -0.322 & -0.329 & -0.331 & -0.332 \
& -0.332 & -0.333 & -0.332 
\\
\hline
{\bf 0.5} & 
-0.283 & -0.267 & -0.284 & -0.308 & -0.323 & -0.330 & -0.332 & -0.333 \
& -0.333 & -0.333 & -0.333 
\\
\hline
\end{tabular}%
}
\caption{Numerical values of Kasner exponent $p_t$, evaluated numerically for the non-zero current.}
\label{fig:TableKasner2}
\end{table}

We do not report here values of $p_x$ and $p_y$ but generally they are different from each other indicating anisotropy of the resulting Kasner space. This apparent anisotropy is due to the oscillating gauge field. In the tested region of constants $\pmtrC_3, \pmtrC_4$ exponent $p_t$ is negative so space is expanding in timelike direction and both $p_x, p_y$ are positive so space is contracting in those directions. It would be interesting to find if there are values which lead to Kasner inversion when $p_t$ becomes positive. 

For a few sets of values of $\pmtrC_3, \pmtrC_4$ the Kasner coefficients are very close to those of a Wick rotated Schwarzschild singularity. For example, when the current vanishes, for $\pmtrC_3=\pmtrC_4=-0.5$ we have
\begin{align}
p_t =-0.333,\qquad
p_x=0.666,\qquad
p_y=0.667.
\end{align}
The same metric arises in the case where we switch off the driving gauge field but we keep non-zero $\pmtrC_3$. This is consistent with the expectation that addition of matter leads to the appearance of a singularity in the bulk. When in addition we turn on $\pmtrC_4$, an anisotropy shows up. Here, because the gauge field does not contribute, an anisotropy in values of $p_x, p_y$ is solely to $\pmtrC_4$ and can be easily controlled. 
However, we found there were no values of $\pmtrC_3, \pmtrC_4$ for which one of the Kasner exponents becomes $1$ (with the others being zero). If that was the a case, there would be no singularity and the geometry would correspond to either a Lifshitz black hole (if $p_t=1$) or a Lifshitz soliton (if $p_x=1$ or $p_y=1$).

\end{appendix}

\printbibliography

\end{document}